\documentclass[draft]{agujournal2019}
\usepackage{url} %this package should fix any errors with URLs in refs.
\usepackage{lmodern}
\usepackage{lineno}

\usepackage{setspace}
\usepackage{amssymb}
\usepackage{amsmath}
\usepackage{tabstackengine}   
\usepackage{graphicx}
\usepackage{graphics}
\usepackage{adjustbox}
\usepackage{float}
\usepackage[arrowdel]{physics}
\usepackage{xcolor}
\usepackage{dsfont}
\usepackage{soul}
\usepackage{enumitem}
\usepackage{soul} 

\makeindex

\newcommand*\patchAmsMathEnvironmentForLineno[1]{
  \expandafter\let\csname old#1\expandafter\endcsname\csname #1\endcsname
  \expandafter\let\csname oldend#1\expandafter\endcsname\csname end#1\endcsname
  \renewenvironment{#1}
  {\linenomath\csname old#1\endcsname}
  {\csname oldend#1\endcsname\endlinenomath}}
  \newcommand*\patchBothAmsMathEnvironmentsForLineno[1]{
  \patchAmsMathEnvironmentForLineno{#1}
  \patchAmsMathEnvironmentForLineno{#1*}}
  \AtBeginDocument{
  \patchBothAmsMathEnvironmentsForLineno{equation}
  \patchBothAmsMathEnvironmentsForLineno{align}
  \patchBothAmsMathEnvironmentsForLineno{flalign}
  \patchBothAmsMathEnvironmentsForLineno{alignat}
  \patchBothAmsMathEnvironmentsForLineno{gather}
  \patchBothAmsMathEnvironmentsForLineno{multline}
}

\draftfalse

\journalname{JGR: Planets}

\begin{document}

%% ------------------------------------------------------------------------ %%
%  Title
%% ------------------------------------------------------------------------ %%

\title{Investigating Barotropic Zonal Flow in Jupiter's Deep Atmosphere using Juno Gravitational Data}

%% ------------------------------------------------------------------------ %%
%
%  AUTHORS AND AFFILIATIONS
%
%% ------------------------------------------------------------------------ %%

\authors{Laura Kulowski\affil{1}, Hao Cao\affil{1}, Rakesh K. Yadav\affil{1}, Jeremy Bloxham\affil{1}}

\affiliation{1}{Department of Earth and Planetary Sciences, Harvard University, Cambridge MA, USA}

\correspondingauthor{Laura Kulowski}{laurakulowski@g.harvard.edu}

%% ------------------------------------------------------------------------ %%
%
%  KEYPOINTS 
%
%% ------------------------------------------------------------------------ %%

% 140 character limit 

\begin{keypoints}

  \item We examine a model in which Jupiter's deep atmospheric zonal flow is barotropic until truncated at depth by a dynamical process 

  \item Most of the Juno gravitational data can be explained by extending the observed winds between $20.9^{\circ}\rm{S}-26.4^{\circ}\rm{N}$ to depths of $\sim 1000$ km
  
  \item The Juno gravitational data can be fully explained by combining these flows with a few broad mid/high latitude jets

  \end{keypoints}

%% ------------------------------------------------------------------------ %%
%
%  ABSTRACT
%
%% ------------------------------------------------------------------------ %%

\newpage 

% 250 words 
\section*{Abstract}
 
\noindent The high-precision Juno gravitational measurements allow us to infer the structure of Jupiter's deep atmospheric zonal flow. Since this inference is nonunique, it is important to explore the space of possible solutions. In this paper, we consider a model in which Jupiter's deep atmospheric zonal flow is barotropic, or invariant along the direction of the rotation axis, until it is truncated at depth by some dynamical process (e.g., Reynolds stress, Lorentz or viscous force). We calculate the density perturbation produced by the $z$-invariant part of the flow using the thermal wind equation and compare the associated odd zonal gravitational harmonics ($J_{3}$, $J_{5}$, $J_{7}$, $J_{9}$) to the Juno-derived values. Most of the antisymmetric gravitational signal measured by Juno can be explained by extending observed winds between $20.9^{\circ}\rm{S}-26.4^{\circ}\rm{N}$ to depths of  $\sim 1000$ km. Because the small-scale features of the mid/high latitude zonal flow may not persist to depth, we allow the zonal flow in this region to differ from the observed surface winds. We find that the Juno odd zonal gravitational harmonics can be fully explained by $\sim 1000$ km deep barotropic zonal flows involving the observed winds between $20.9^{\circ}\rm{S}-26.4^{\circ}$N and a few broad mid/high latitude jets.

%% ------------------------------------------------------------------------ %%
%
%  PLAIN LANGUAGE SUMMARY 
%
%% ------------------------------------------------------------------------ %%

\section*{Plain Language Summary}

\noindent One of Jupiter's most recognizable features is its surface pattern of zonal (east-west) winds. Since large-scale flows are expected to produce a signature in the planet's gravitational field, we can use the Juno gravitational data to test different zonal flow models. In this paper, we test a model in which the zonal flow extends into the interior along the direction of the planetary rotation axis without decay until it is truncated at depth by some dynamical process. The model remains agnostic on the exact truncation mechanism, although possibilities include interaction with a stably stratified layer or the magnetic field. We find that most of the dynamical gravitational field measured by Juno can be explained by extending the observed surface winds between $20.9^{\circ}$S-$26.4^{\circ}$N to depths of $\sim1000$ km. The Juno data can be fully explained when these flows are combined with a few broad mid/high latitude jets with peak amplitudes of about $10$ m s$^{-1}$.

%% ------------------------------------------------------------------------ %%
%
%  MAIN TEXT
%
%% ------------------------------------------------------------------------ %%

\newpage

\section{Introduction}
\label{sec:intro}
\noindent 

The zonal winds within Jupiter's atmosphere give rise to the planet's familiar banded appearance. For almost fifty years, researchers have debated whether the observed winds represent shallow weather-layer dynamics or deep-seated convection \cite{cuzzi69, showman_2007,busse76}. The recent high-precision Juno gravitational measurements \cite{iess18, durante20} allow us to infer possible deep atmospheric zonal flow profiles, making it possible to  distinguish between these two endpoint hypotheses. Previous studies have shown that the Juno gravitational measurements are consistent with baroclinic zonal flows that extend about $3000$ km deep \cite{kaspi18, kaspi20, kong18, duer20}. 

Determining the deep atmospheric zonal flow profile that corresponds to the Juno gravitational data is a nonunique problem  \cite{kaspi18, kaspi20, stevenson20}. The density perturbations associated with a zonal flow profile are typically calculated using the thermal (gravitational) wind equation \cite{kaspi13, zhang15, hao17, wicht19}. The same density perturbation, and thus the same gravitational signal, can be generated by slow zonal flow in regions where the background density is large and fast zonal flow in regions where the background density is small. Given the inherent nonuniqueness of the gravitational inference, it is important to explore the space of possible zonal flow profiles. One meaningful way to do this is to test physically motivated models of deep atmospheric zonal flow against the Juno data. 

\begin{figure}[t]
  \begin{adjustbox}{center}
      \includegraphics[width=37pc]{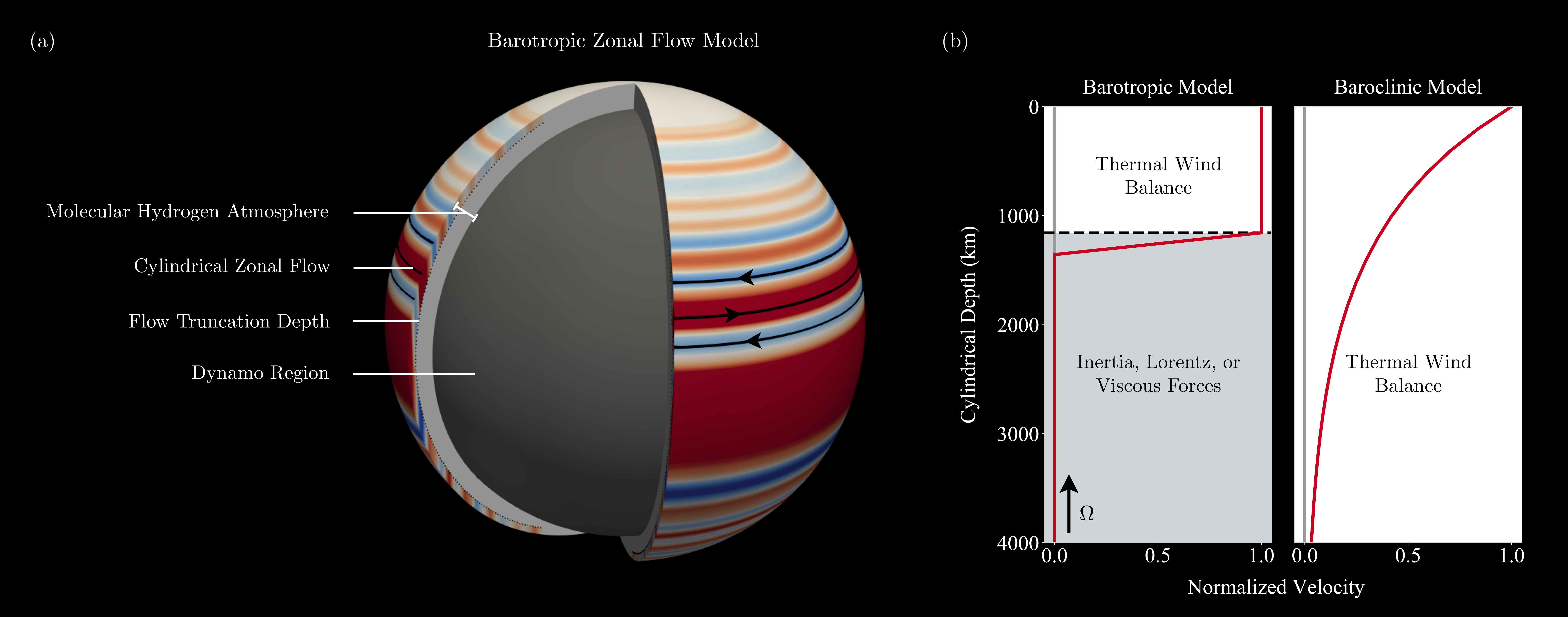}
    \end{adjustbox}
    \caption{(a) Example barotropic zonal flow profile where the observed surface zonal winds \cite{tollefson17} extend into the interior along the $z$-direction until they are truncated at depth by some dynamical process (dashed lines). Prograde/retrograde jets are indicated in red/blue with a color scale that ranges between $\pm 50$ m s$^{-1}$. (b) Comparison of the dynamical balances for our barotropic model and the baroclinic models considered in previous studies. The barotropic zonal flow satisfies the thermal wind equation down to the truncation depth (dashed line), at which point inertial, Lorentz, or viscous forces act to reduce the flow speed. In contrast, the baroclinic zonal flow satisfies the thermal wind equation throughout Jupiter's atmosphere and the flow decays with depth due to baroclinic density variations.} 
      \label{fig:bt}
    \end{figure}

In this paper, we test a model in which Jupiter's zonal flow is barotropic and satisfies the Taylor-Proudman constraint (i.e., is invariant along the rotation axis) until it is truncated at depth by some dynamical process (Fig.~\ref{fig:bt}a). We are motivated to examine this model for two reasons. First, in Jupiter's deep atmosphere where the effects of viscosity and inertia can be neglected, the Taylor-Proudman constraint describes the leading-order dynamics \cite{taylor23}. Second, barotropic zonal flows are commonly featured in models explaining the formation of Jupiter's zonal winds through deep-seated convection. \citeA{busse76} originally proposed that Jupiter's observed surface zonal winds are the surface manifestations of deep zonal flow generated by tilted cylindrical convection columns. Since then, three-dimensional numerical simulations of convection in spherical shells have found that the flow organizes itself into broad and deep zonal jets whose axes are approximately parallel to the rotation axis \cite{christensen01,aurnou01,heimpel05,jones09, gastine14z, yadav20_hex}.

Previous studies have focused on baroclinic models and considered zonal flow profiles that decay rapidly in the upper atmosphere \cite{kaspi18, kaspi20, kong18, duer20, galanti21} or are $z$-invariant until they decay baroclinically at depth \cite{galanti20, dietrich21}. Our barotropic model is fundamentally different from these baroclinic models because it makes no assumption as to what causes the winds to be truncated at depth (Fig.~\ref{fig:bt}b). The barotropic zonal flow satisfies the thermal wind equation down to the truncation depth, at which point other forces (e.g., Reynolds stress, Lorentz or viscous force) enter the dynamical balance and reduce the flow speed to much smaller values. Thus, there are a range of different physical processes that could act to truncate the flow. In baroclinic models, the flow satisfies the thermal wind equation throughout the deep atmosphere and decays with depth due to baroclinic density perturbations. The truncation mechanism is an inherent assumption of the baroclinic model, and no additional forces need to be considered.

This paper is organized as follows. In Section~\ref{sec:theory}, we review the dynamical balances governing barotropic zonal flows and show how to calculate the flow-induced density perturbations and gravitational signals. In Section~\ref{sec:results}, we compare the odd zonal gravitational harmonics ($J_{3}$, $J_{5}$, $J_{7}$, $J_{9}$) produced by different barotropic zonal flow profiles to the values measured by Juno. Finally, in Section~\ref{sec:end}, we summarize our results and discuss their implications.

  \section{Theory}
  \label{sec:theory}

  Fluid motions within Jupiter's deep atmosphere are governed by the momentum equation, which in a fixed rotating reference frame is given by  
  \begin{align}
    \underbrace{\vphantom{\frac{X}{\gamma}}  \frac{\partial \vb{u}}{\partial t}  + \vb{u} \cdot \nabla \vb{u} } _\text{Inertial} 
    + \underbrace{\vphantom{\frac{X}{\gamma}} 2\vb{\Omega} \times \vb{u} } _\text{Coriolis}
    = \underbrace{\vphantom{\frac{X}{\gamma}} -\frac{1}{\rho} \nabla p }_{\begin{subarray}{c}\text{Pressure} \\ \text{Gradient}\end{subarray}} 
    \hspace{0.09cm} \underbrace{\vphantom{\frac{X}{\gamma}} - \nabla U }_{\begin{subarray}{c}\text{Effective} \\ \text{Gravity}\end{subarray}} %_\text{Effective Gravity}
    + \underbrace{\vphantom{\frac{X}{\gamma}}  \frac{1}{\mu_{0} \rho} \qty(\nabla \times \vb{B}) \times \vb{B} } _\text{Lorentz}
    - \underbrace{\vphantom{\frac{X}{\gamma}}  \frac{1}{\rho}\qty(\nabla \cdot \vb{T}_{visc}) } _\text{Viscous} 
    \label{eq:mom_full}
  \end{align}
  where $\vb{u}$ is the velocity in the rotating frame, $\vb{\Omega} = \Omega_{0} \vb{\hat{z}}$ is the rotational angular velocity of the planet, $\rho$ is the density, $p$ is the pressure, $U$ is the effective gravitational potential (i.e., the sum of the gravitational and centrifugal potentials and is related to the effective gravity by $\vb{g_{\rm{eff}}} = -\nabla U$), $\mu_{0}$ is the permeability of free space, $\vb{B}$ is the magnetic field, and $\vb{T}_{visc}$ is the viscous stress tensor. The physical forces involved in this equation are indicated below the relevant terms. Due to Jupiter's rapid rotation, Coriolis accelerations are expected to dominate over inertial accelerations and viscous forces on large length scales. This leads to the common practice of neglecting the inertial and viscous terms in Eq.~\ref{eq:mom_full} \cite{kaspi09, kaspi13,zhang15,hao17}. Since the upper $\sim2000$ km of the deep atmosphere is electrically nonconducting \cite{french12}, the contribution from the Lorentz force may also be neglected. Removing the inertial, viscous, and Lorentz forces, Eq.~\ref{eq:mom_full} reduces to      
  \begin{align}
    2\vb{\Omega} \times \vb{u} = -\frac{1}{\rho} \nabla p + \vb{g_{\rm{eff}}} .
    \label{eq:mom2}
  \end{align}
  We can gain information about the structure of the barotropic flow (Section~\ref{sec:flow_structure}) and its associated density perturbation and gravitational signal (Section~\ref{sec:gravity}) by examining the curl of Eq.~\ref{eq:mom2}.

  \subsection{Flow Structure}
  \label{sec:flow_structure}

  Taking the curl of Eq.~\ref{eq:mom2} directly yields 
  \begin{align}
    \qty(2 \boldsymbol{\Omega} \cdot \nabla) \vb{u} - 2 \boldsymbol{\Omega} \qty(\nabla \cdot \vb{u}) = -\frac{1}{\rho^{2}}\qty(\nabla \rho \times \nabla p).
    \label{eq:vort1} 
  \end{align}
  The zonal component of Eq.~\ref{eq:vort1} is given by 
  \begin{align}
    2\Omega_{0}\frac{\partial u_{\phi}}{\partial z} = \frac{1}{\rho^{2}}[\nabla \rho \times \nabla p ]_{\phi} . 
    \label{eq:vort1_phi}
  \end{align}
    For our barotropic model, where $\nabla \rho \times \nabla p = 0$, Eq.~\ref{eq:vort1_phi} tells us that the zonal flow obeys the Taylor-Proudman theorem and is invariant along the direction of the axis of rotation (i.e., $\partial u_{\phi} / \partial z = 0$). If the planet remains barotropic almost everywhere, in order for the flow to be truncated at depth, other terms from the momentum equation (e.g., Reynolds stress, Lorentz or viscous force) must reenter the momentum equation and balance the vertical shear of the flow.

  \subsection{Gravity Calculation}
  \label{sec:gravity}

  Alternatively, we can multiply Eq.~\ref{eq:mom2} by the density before taking the curl to obtain 
  \begin{align}
    2\qty(\vb{\Omega} \cdot \nabla) \rho \vb{u} = - \nabla \rho \times \vb{g_{\rm{eff}}}.
    \label{eq:vort2}
  \end{align}
  The zonal component of Eq.~\ref{eq:vort2} is then given by 
  \begin{align}
    2\Omega_{0}\frac{\partial}{\partial z}\qty(\rho u_{\phi}) = - \qty[\nabla \rho \times \vb{g_{\rm{eff}}}]_{\phi}.
    \label{eq:vort2_phi}
  \end{align}
  Although barotropic zonal flow has zero vertical wind shear, it has nonzero vertical momentum shear due to Jupiter's rapidly varying background density (i.e., $\partial\qty(\rho u_{\phi}) /\partial z \ne 0$). Since the left-hand side of Eq.~\ref{eq:vort2_phi} is nonzero for $z$-invariant zonal flows within Jupiter's deep atmosphere, we can relate the barotropic zonal flow to the planet's density distribution. 
  
  To calculate the flow-induced density perturbation, we separate the density and effective gravity in Eq.~\ref{eq:vort2_phi} into static components ($\rho_{0}$, $\vb{g_{\rm{eff}_0}}$), which are associated with Jupiter's hydrostatic background state, and dynamic components ($\rho '$, $\vb{g_{\rm{eff}}}'$), which are associated with the flow, so that 
\begin{align}
  \rho\qty(r, \theta) &=  \rho_{0}\qty(r) + \rho ' \qty(r, \theta) 
  \label{eq:rho_exp} \\
  \vb{g_{\rm{eff}}}\qty(r, \theta) &=  \vb{g_{\rm{eff}_{0}}}\qty(r) + \vb{g_{\rm{eff}}}'\qty(r, \theta).
\end{align}
Substituting these expansions into Eq.~\ref{eq:vort2_phi} gives us the leading order balance 
\begin{align}
  2\Omega_{0}   \frac{\partial}{\partial z}\qty(\rho_{0} u_{\phi}) = -\qty[\nabla \rho ' \times \vb{g_{\rm{eff}_{0}}} + \nabla \rho_{0} \times \vb{g_{\rm{eff}}}']_{\phi}. 
  \label{eq:tgwe}
\end{align}
In Eq.~\ref{eq:tgwe}, the term, $\nabla \rho_{0} \times \vb{g_{\rm{eff}}}'$ is sometimes referred to as the dynamic self-gravity \cite{zhang15,wicht19}. In the context of studying zonal flows in a gas giant atmosphere where the background density rapidly increases with depth, the importance of this term is a subject of debate. While some studies argue that the dynamic self-gravitation is on the same order as $\nabla \rho ' \times \vb{g_{\rm{eff}_{0}}}$ when assuming a rigid spherical outer boundary \cite{zhang15, kong18}, others find that this term has a small impact \cite{eli17, kaspi18} or is counterbalanced by the effects of rotational deformation \cite{hao17}. In this paper, we neglect the contribution from the dynamic self-gravitation term. We will show later in Section \ref{sec:compare} that neglecting this term does not change our conclusions. Without the dynamic self-gravity, Eq.~\ref{eq:tgwe} reduces to the thermal wind equation (as defined in \citeNP{kaspi10}) 
\begin{align} 
   2\Omega_{0}   \frac{\partial}{\partial z}\qty(\rho_{0} u_{\phi}) = -\qty[\nabla \rho ' \times \vb{g_{\rm{eff}_{0}}} ]_{\phi}.
  \label{eq:twe}
\end{align}
For simplicity, we assume that Jupiter has a spherical background shape. We can solve for the dynamical density perturbation by integrating Eq.~\ref{eq:twe} along the $\theta$-direction to obtain 
\begin{align}
  \rho'\qty(r, \theta) = - \frac{2\Omega_{0}r}{g_{0}\qty(r)} \frac{\partial \rho_{0}\qty(r)}{\partial z} \int_{0}^{\theta ' = \theta} u_{\phi}\qty(r, \theta') d\theta ' + \rho_{c}'\qty(r),
  \label{eq:rho_prime}
\end{align}
where $g_{0}\qty(r)$ is the gravitational acceleration and $\rho_{c}'\qty(r)$ is a constant function of integration. Eq.~\ref{eq:rho_prime} establishes the relationship between the barotropic zonal flow and its associated density perturbation. 

The flow-induced density perturbation will leave a signature in the planet's gravitational field. We can calculate the zonal gravitational harmonics produced by the barotropic zonal flow by evaluating 
\begin{align}
  \Delta J_{n} = -\frac{1}{Ma^{n}}\int_{\mathcal{V}} \rho'\qty(r', \theta ') P_{n}\qty(\cos\theta  ' ) d^{3}\vb{r}',  
  \label{eq:jns}
\end{align}
where $M = 1.89819 \times 10^{27}$ kg is Jupiter's mass, $a = 71, 492$ is the equatorial radius at 1 bar, $n$ is the degree, $\mathcal{V}$ is the volume of the planet, and $P_{n}$ are the Legendre polynomials.

\section{Results}
\label{sec:results}

\subsection{Barotropic Zonal Flows that Match Jupiter's Surface Zonal Winds}
\label{sec:obs} 

\begin{figure}[t]
  \centering
  \begin{adjustbox}{center}
  \includegraphics[width=35pc]{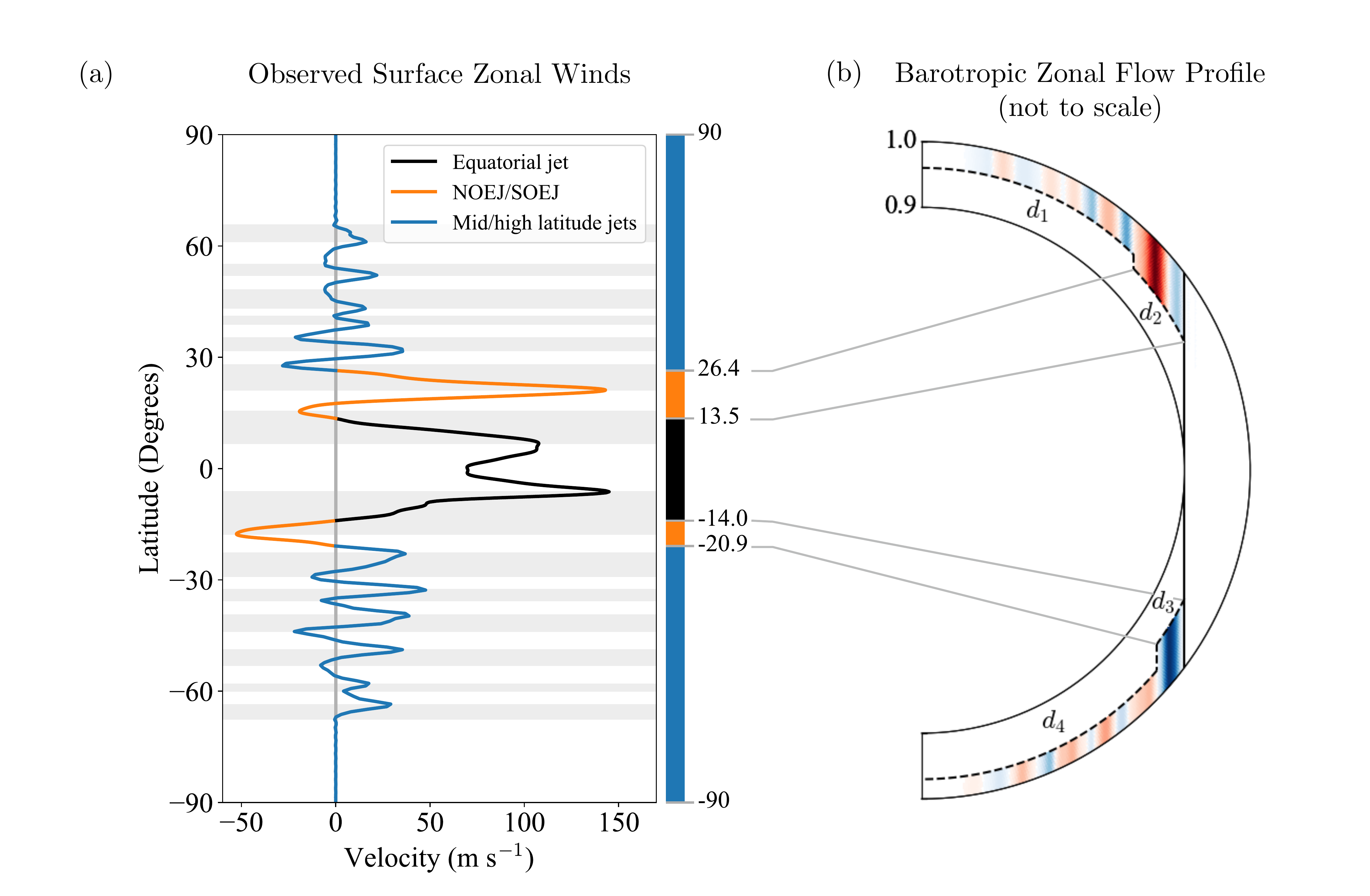}
\end{adjustbox}
\caption{(a) We separate the observed surface zonal winds  \cite{tollefson17} into three latitudinal regions based on morphology: the equatorial jet (black), the northern and southern off-equatorial jets (NOEJ/SOEJ, orange), and the mid/high latitude jets (blue). Gray (white) bands indicate belts (zones). (b) Barotropic zonal flow profiles are generated by extending the NOEJ, SOEJ, and mid/high latitude jets into the interior along the $z$-direction to radial depths $d_{i}$ (dashed lines). The barotropic zonal flow profile shown in panel (b) is not to scale.}
  \label{fig:separation}
\end{figure}

We begin our analysis by assuming that the observed surface zonal winds \cite{tollefson17} extend into the interior barotropically until they are truncated at depth by some unspecified dynamical process. Motivated by differences in morphology, we divide Jupiter's observed surface winds into three distinct regions: first, we identify the equatorial jet ($14.0^{\circ}$S--$13.5^{\circ}$N in planetocentric latitude), which is predominately symmetric with respect to the equator; second, the northern off-equatorial jets (NOEJ, $13.5^{\circ}$--$26.4^{\circ}$N) and the southern off-equatorial jet (SOEJ, $14.0^{\circ}$--$20.9^{\circ}$S), which are strongly antisymmetric about the equator; and third, the mid/high latitude jets ($26.4^{\circ}$--$90^{\circ}$N and $20.9^{\circ}$--$90^{\circ}$S), which are dominated by small-scale features (Fig.~\ref{fig:separation}a). Since we are interested in antisymmetric zonal flows that produce odd zonal gravitational harmonics, we neglect the equatorial jet from our analysis (see \ref{sec:equ_jet} for justification). In the other regions, we allow the flow depth to vary. We define $d_{1}$ as radial depth of the northern mid/high latitude jets, $d_{2}$ as the radial depth of the NOEJ, $d_{3}$ as the radial depth of the SOEJ, and $d_{4}$ as the radial depth of the southern mid/high latitude jets (Fig.~\ref{fig:separation}b).

\begin{figure}[t]
  \centering
  \begin{adjustbox}{center}
  \includegraphics[width = 35pc]{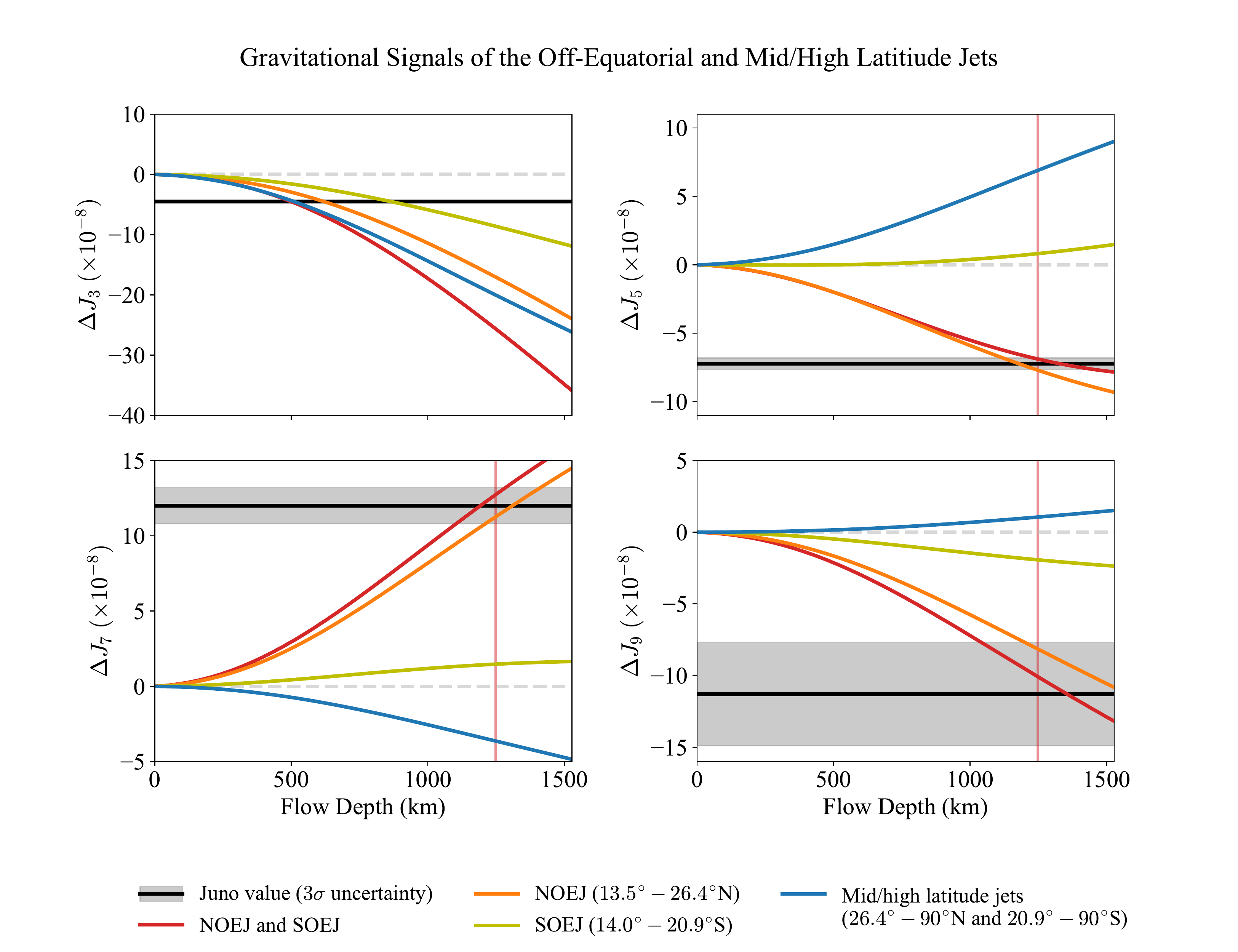}
\end{adjustbox}
\caption{Odd zonal gravitational harmonics as a function of flow depth for barotropic zonal flows within the off-equatorial and mid/high latitude jets (see Fig.~\ref{fig:separation} for definitions). The Juno measured values and their $3\sigma$ uncertainties \cite{durante20} are shown in black and gray, respectively. When the NOEJ and SOEJ both extend $1244-1253$ km deep (red shaded bars), they produce $J_{5}$, $J_{7}$, and $J_{9}$ values that are consistent with the Juno measurements.}
  \label{fig:lat_regions}
\end{figure}

\begin{figure}[t]
    \begin{adjustbox}{center}
    \includegraphics[width=38pc]{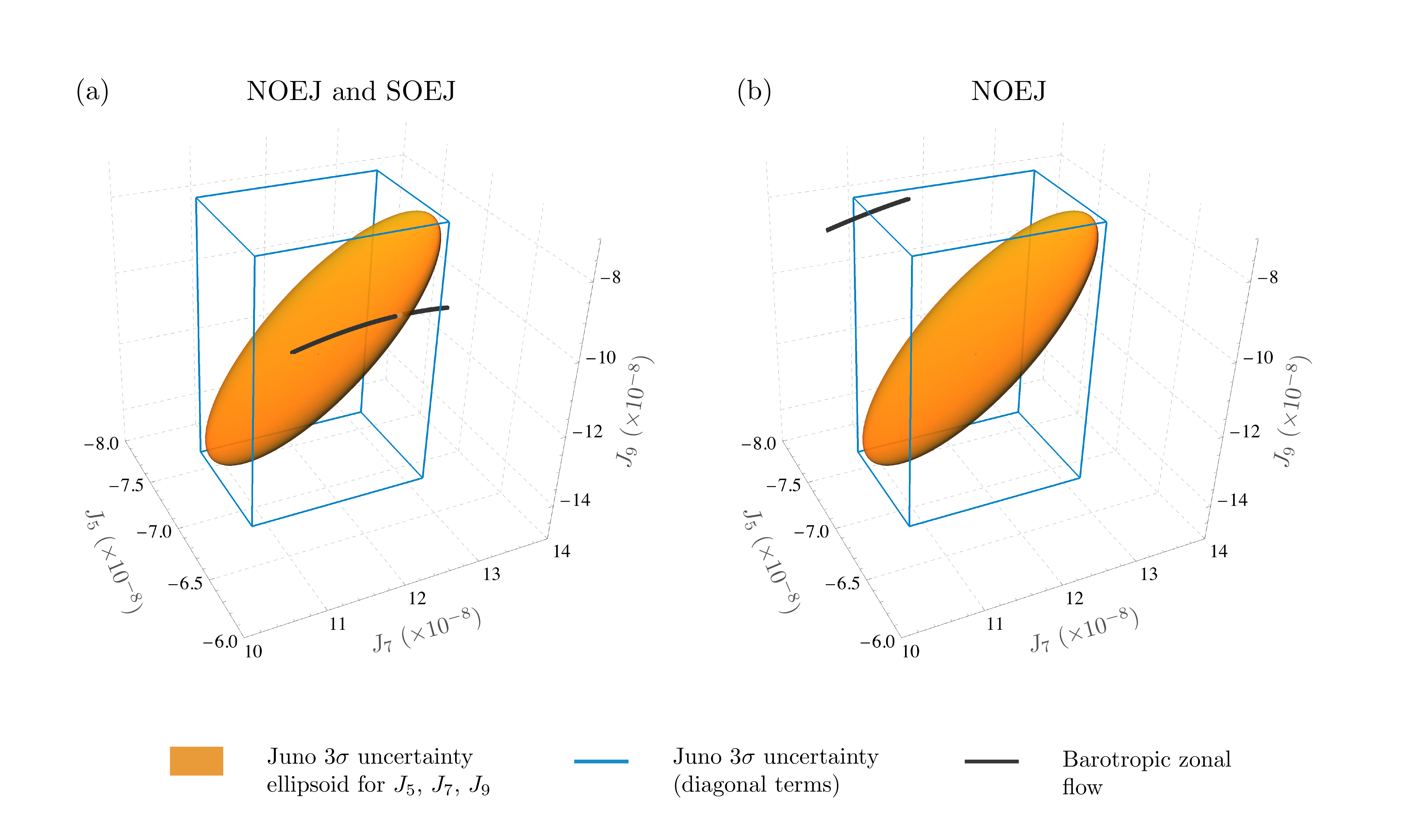}
        \end{adjustbox}
    \caption{Three-dimensional visualization of the $J_{5}$, $J_{7}$, and $J_{9}$ values produced by the (a) the NOEJ and SOEJ and (b) the NOEJ (gray). Flows involving the NOEJ and SOEJ intersect the $3\sigma$ uncertainty ellipsoid associated with the Juno measurements (orange), while flows involving only the NOEJ do not. The $3\sigma$ uncertainty associated with each harmonic (i.e., the diagonal terms in the error covariance matrix) are marked by blue boxes.}
  \label{fig:lat_regions_3d}
  \end{figure}

We calculate the odd zonal gravitational harmonics as a function of flow depth for individual latitudinal regions (Fig.~\ref{fig:lat_regions}). By symmetrically extending the NOEJ and SOEJ into the interior, we consider the set of flows described by 
\begin{align}
  \mathcal{D}_{nsjets} = \{ \qty(d_{1}, \: d_{2}, \: d_{3}, \: d_{4}) \mid \: d_{1} = d_{4} = 0, \: d_{2} = d_{3}, \: 0 \le d_{2}, d_{3} \le 1530 \:\rm{km} \}.
  \label{eq:set_nsjets}
\end{align}
When the NOEJ and SOEJ extend $1244-1253$ km deep, they can produce $J_{5}$, $J_{7}$, and $J_{9}$ values that fall within the $3\sigma$ uncertainty of the Juno measurements (Fig.~\ref{fig:lat_regions}, red curves; Fig.~\ref{fig:lat_regions_3d}a). For these depths, the value of $J_{3}$ is about one order of magnitude larger than the Juno value. The gravitational contribution from zonal flows within Jupiter's dynamo region, which typically produces $J_{3}$ values of between $1.69 - 3.22 \times 10^{-8}$, is not large enough to correct the deep atmospheric values \cite{kulowski20}. The NOEJ and SOEJ can therefore explain most of the asymmetric gravitational field observed by Juno, but struggle to produce the measured $J_{3}$ value.  

If we examine the individual gravitational contributions of the NOEJ and SOEJ (i.e., either $d_{2} = 0$ or $d_{3} = 0$ in Eq.~\ref{eq:set_nsjets}), we find that the NOEJ is responsible for most of the signal (Fig.~\ref{fig:lat_regions}, orange and yellow curves). The odd gravitational harmonics produced by the NOEJ, however, do not pass through the Juno $3\sigma$ ellipsoid (Fig.~\ref{fig:lat_regions_3d}b). This implies that the NOEJ and SOEJ are both required to match the Juno $J_{5}$, $J_{7}$, and $J_{9}$ values.  

Moving on to the mid/high latitude jets, we consider the set of flows described by 
\begin{align}
  \mathcal{D}_{mid/high \:lats} = \{\qty(d_{1}, \: d_{2}, \: d_{3}, \: d_{4}) \mid \: d_{2} = d_{3} = 0, \: d_{1} = d_{4}, \: 0 \le d_{1}, d_{4} \le 1530 \:\rm{km}\}.
  \label{eq:set_midlats}
\end{align}
For depths of about $1000$ km, the mid/high latitude flows are capable of producing odd zonal gravitational harmonics that have similar magnitudes to the Juno-derived values (Fig.~\ref{fig:lat_regions}, blue curves). While the $J_{3}$ value has the same sign as the Juno value, the $J_{5}$, $J_{7}$, and $J_{9}$ values have opposite signs. The mid/high latitudes alone cannot account for the observed antisymmetric gravitational field. Rather, when these flows are added to the northern and southern jets, they act to increase the magnitude of $J_{3}$ and decrease the magnitudes of $J_{5}$, $J_{7}$, and $J_{9}$.    

Our analysis of barotropic zonal flows that match the observed surface zonal winds has shown that NOEJ and SOEJ play an important role in explaining the Juno gravitational data. On their own, these jets can reproduce most features of Jupiter's observed antisymmetric gravitational field.

\subsection{Barotropic Zonal Flows that Differ from Jupiter's Surface Zonal Winds}
\label{sec:smooth} 

In Section \ref{sec:obs}, we assumed that Jupiter's observed surface zonal winds were barotropic. However, it is possible that the weaker, smaller scale features in the mid/high latitudes do not extend into the interior undisturbed. In this section, we explore the possibility that the mid/high latitude zonal flows at depth consist of a few broad jets. We investigate whether smooth mid/high latitude flows can be combined with the NOEJ and SOEJ to produce gravitational signals that are consistent with the Juno data.

\begin{figure}[t]
  \centering
  \begin{adjustbox}{center}
  \includegraphics[width = 45pc]{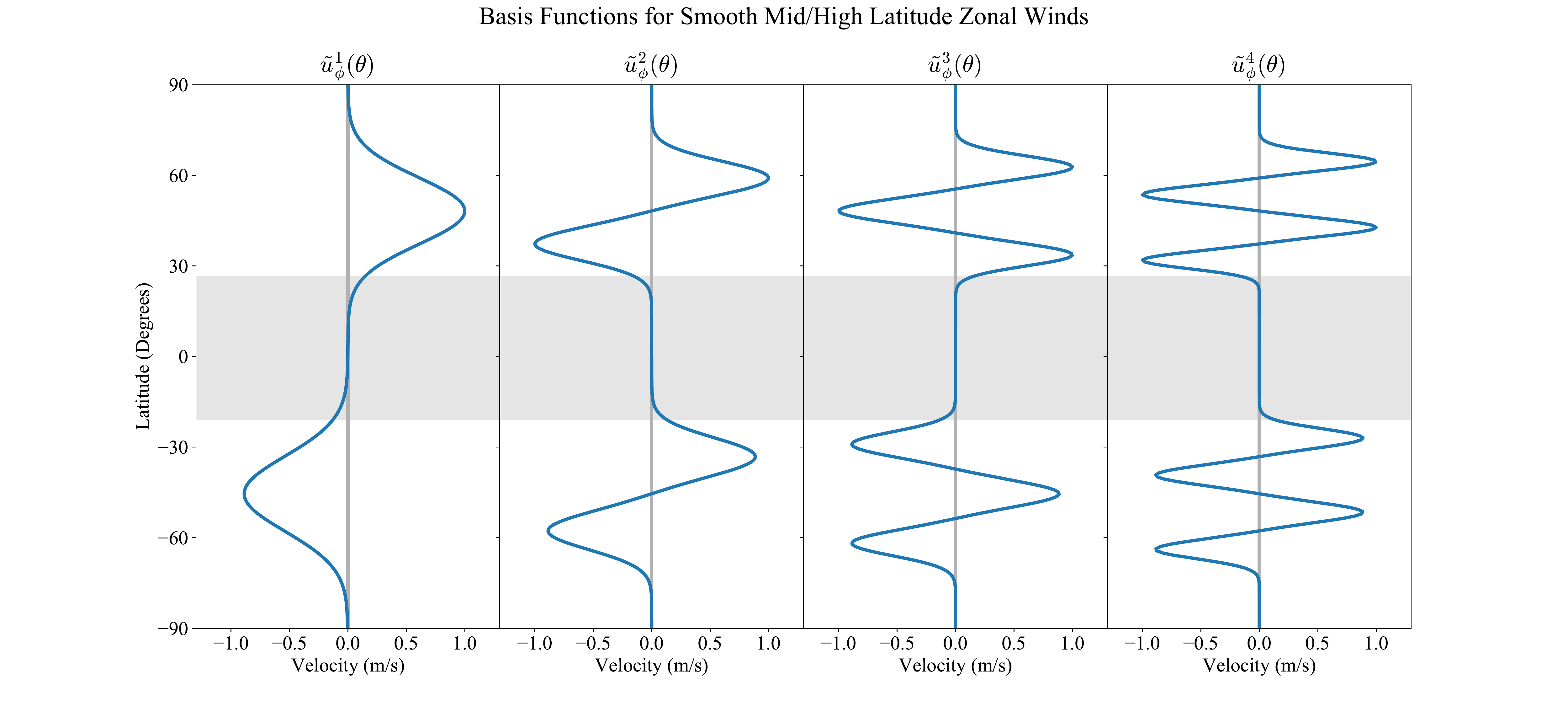}
\end{adjustbox}
\caption{Basis functions, $\tilde{u}_{\phi}^{i}$, defined at Jupiter's surface that specify the smooth zonal winds in the mid/high latitudes. These basis functions contain one to four jets in each hemisphere. To generate the interior zonal flow profile, we extend each basis function into the interior barotropically to the same truncation depth and then linearly combine the four flow fields. The gray shaded bands indicate the latitudes occupied by the equatorial jet, NOEJ, and SOEJ.}
  \label{fig:basis}
\end{figure}

The first step in this investigation is to construct smooth mid/high latitude barotropic zonal flows. To do this, we define four basis functions, $\tilde{u}_{\phi}^{i}\qty(\theta)$, for the surface zonal winds in the mid/high latitudes (Fig.~\ref{fig:basis}). These basis functions contain between one and four jets in each hemisphere. The observed surface winds, in contrast, contain about ten jets in each hemisphere. We extend each basis function into the interior barotropically to the same truncation depth and then linearly combine the four flow fields to obtain the zonal flow profile for the mid/high latitudes. 

We now examine the gravitational signal that would be produced by combining the NOEJ and SOEJ with the smooth mid/high latitude zonal flows. We assume that the barotropic zonal flow is truncated at a single depth between $1000-1530$ km. For these truncation depths, the NOEJ and SOEJ produce $J_{5}$, $J_{7}$, and $J_{9}$ values that are close to the Juno values, but generate $J_{3}$ values that are too large in magnitude (Fig.~\ref{fig:lat_regions}). The smooth mid/high latitude flows are primarily responsible for shifting the $J_{3}$ value associated with the NOEJ and SOEJ to the Juno measured value while providing small corrections to the $J_{5}$, $J_{7}$, and $J_{9}$ values. To solve for the smooth flows that would accomplish this, we compute the odd zonal gravitational harmonics produced by each basis flow. Since the zonal flow is directly proportional to the zonal gravitational harmonics, we can rescale the flows and their associated gravitational contributions so that they produce the required odd zonal gravitational harmonics. This amounts to solving the system of equations 

\begin{figure}[t]
  \begin{adjustbox}{center}
  \includegraphics[width = 40pc]{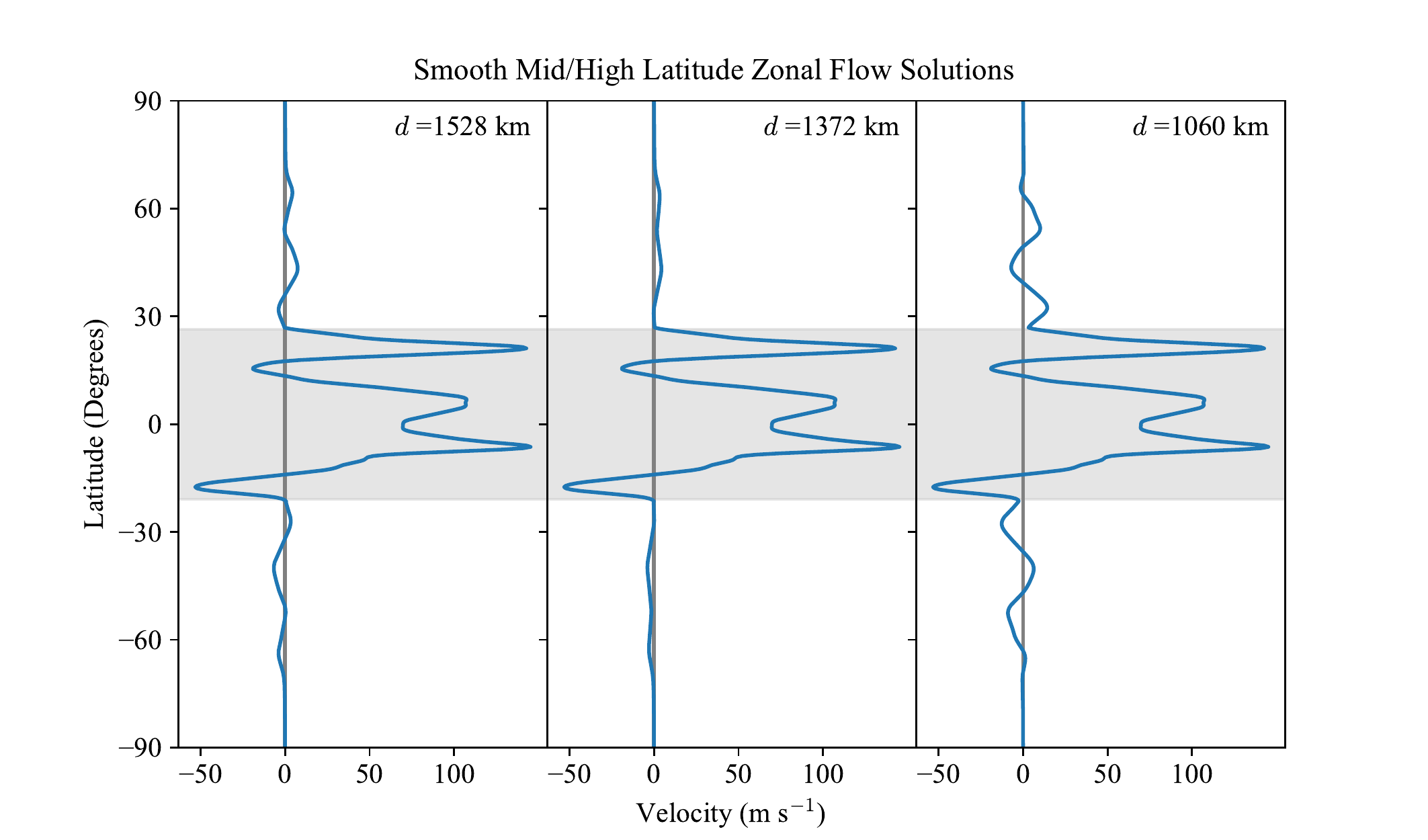}
\end{adjustbox}
  \caption{Surface wind profiles with smooth mid/high latitude zonal flow. The truncation depth of the flow is indicated in the top right of each panel and ranges between $1000-1530$ km ($5-15$ kbar). The gray shaded bands indicate the regions occupied by the equatorial jet, NOEJ, and SOEJ.}
  \label{fig:smooth}
\end{figure}

\setstacktabbedgap{2ex}
\setstackgap{L}{2.0\baselineskip}

\begin{equation}
  \parenMatrixstack{ 
  J_{3}\qty(\tilde{u}_{\phi}^{1}) & J_{3}\qty(\tilde{u}_{\phi}^{2}) & J_{3}\qty(\tilde{u}_{\phi}^{3}) & J_{3}\qty(\tilde{u}_{\phi}^{4})\\
    J_{5}\qty(\tilde{u}_{\phi}^{1}) & J_{5}\qty(\tilde{u}_{\phi}^{2}) & J_{5}\qty(\tilde{u}_{\phi}^{3}) & J_{5}\qty(\tilde{u}_{\phi}^{4})\\
    J_{7}\qty(\tilde{u}_{\phi}^{1}) & J_{7}\qty(\tilde{u}_{\phi}^{2}) & J_{7}\qty(\tilde{u}_{\phi}^{3}) & J_{7}\qty(\tilde{u}_{\phi}^{4})\\
    J_{9}\qty(\tilde{u}_{\phi}^{1}) & J_{9}\qty(\tilde{u}_{\phi}^{2}) & J_{9}\qty(\tilde{u}_{\phi}^{3}) & J_{9}\qty(\tilde{u}_{\phi}^{4})
  }
  \parenVectorstack{
    \alpha_{1} \vphantom{ J_{3}\qty(\tilde{u}_{\phi}^{1}) }\\
    \alpha_{2} \vphantom{ J_{3}\qty(\tilde{u}_{\phi}^{1}) }\\
    \alpha_{3} \vphantom{ J_{3}\qty(\tilde{u}_{\phi}^{1}) }\\ 
    \alpha_{4} \vphantom{ J_{3}\qty(\tilde{u}_{\phi}^{1}) }
  } = 
  \parenVectorstack{
  \Delta J_{3}^{\rm{smooth}} \vphantom{ J_{3}\qty(\tilde{u}_{\phi}^{1}) }\\ 
  \Delta J_{5}^{\rm{smooth}} \vphantom{ J_{3}\qty(\tilde{u}_{\phi}^{1}) }\\ 
  \Delta J_{7}^{\rm{smooth}}  \vphantom{ J_{3}\qty(\tilde{u}_{\phi}^{1}) }\\
  \Delta J_{9}^{\rm{smooth}}  \vphantom{ J_{3}\qty(\tilde{u}_{\phi}^{1}) }
   }
  \label{eq:jeremy}
\end{equation}
where $J_{n}\qty(\tilde{u}_{\phi}^{i})$ indicates the zonal gravitational harmonic produced by basis flow $\tilde{u}_{\phi}^{i}$, $\alpha_{i}$ is the weight for basis flow $\tilde{u}_{\phi}^{i}$, and $\Delta J_{n}^{\rm{smooth}} = J_{n}^{\rm{Juno}} - J_{n}^{\rm{nsjets}}$ is the difference between the $J_{n}$ values measured by Juno and those produced by the NOEJ and SOEJ. The smooth flow solution is given by 
\begin{align}
  \tilde{u}_{\phi}\qty(r,\theta) = \sum_{i=1}^{4} \alpha_{i} \tilde{u}_{\phi}^{i}\qty(r, \theta). 
  \label{eq:u_smooth}
\end{align}

For each truncation depth between $1000-1530$ km, we find a smooth flow solution that can be combined with the NOEJ and SOEJ to produce odd gravitational harmonics that match Juno measurements. The morphology of the smooth flows varies with truncation depth. Three representative examples are shown in Fig.~\ref{fig:smooth}. Some of the solutions (Fig.~\ref{fig:smooth}a, c) consist of $3-4$ alternating band of zonal flow in each hemisphere, while others (Fig.~\ref{fig:smooth}b) contain $2$ prograde jets in each hemisphere. The peak amplitude of the smooth flows is on the order of $\sim 10$ m s$^{-1}$, which is slower than the $10-65$ m s$^{-1}$ speeds of the observed mid/high latitude zonal winds. 

In this section, we demonstrated that Jupiter's observed gravitational field is consistent with $\sim 1000$ km deep barotropic zonal flows involving the NOEJ, SOEJ, and a few broad, slow jets in the mid/high latitudes. While we have focused on one family of solutions, other smooth mid/high latitude flows could be found by considering different basis flows for the mid/high latitudes (i.e., Fig.~\ref{fig:basis}) or by requiring the smooth flows to satisfy a different set of constraints (i.e., Eq.~\ref{eq:jeremy}).

\subsection{Truncation of the Barotropic Zonal Flow}

In Sections~\ref{sec:obs} and \ref{sec:smooth}, we calculated the gravitational signal produced by the $z$-invariant part of the zonal flow. At the truncation depth, there will be an additional density perturbation associated with the decay of the flow. Determining the form of this density perturbation requires further modeling efforts. If, for example, the barotropic zonal flow experiences magnetic breaking by the Lorentz force, one would need to simulate the interaction between the magnetic field and zonal flow in order to determine how the flow decays with depth. Given the flow decay profile, the dynamical density perturbation could be calculated by solving a modified thermal wind equation that includes the curl of the Lorentz force.   

\begin{figure}[t]
  \begin{adjustbox}{center}
    \includegraphics[width=42pc]{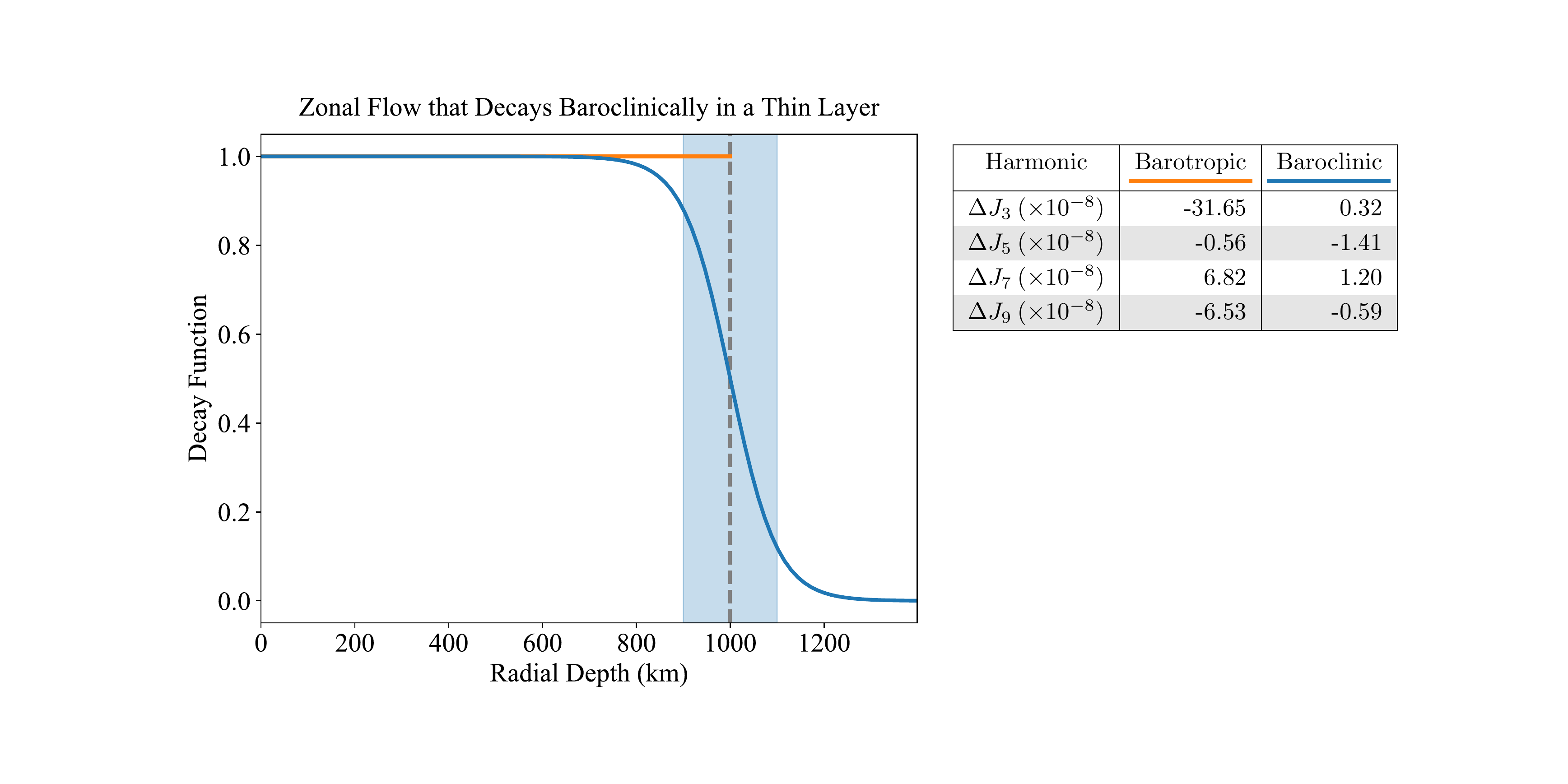}
  \end{adjustbox}
  \caption{Radial decay functions applied to a zonal wind profile where the observed surface winds extend into the interior along the $z$-direction without decay. In the barotropic case (orange), the zonal flow extends $1000$ km deep. In the baroclinic case (blue), the zonal flow is constant until it decays rapidly over a $200$ km thick layer (blue shaded region). The odd zonal gravitational harmonics produced by the barotropic and baroclinic zonal flows are listed in the table.}
  \label{fig:thin_layer}
\end{figure}

The density perturbation associated with the decay of the barotropic zonal flow may contribute to the odd zonal gravitational harmonics. To get a sense of this contribution, we consider a baroclinic zonal flow profile where the observed surface zonal winds are constant along the $z$-direction until they encounter a thin layer where the flow amplitudes rapidly decay to zero. We describe the radial decay of the flow using a hyperbolic tangent function  
\begin{align}
  f\qty(r) = \frac{\tanh\qty(-\frac{a - H - r}{\Delta H}) + 1}{\tanh\qty(\frac{H}{\Delta H} + 1)}
  \label{eq:tanh}
\end{align}
where $a$ is Jupiter's radius, $H$ is the depth at which the surface flow amplitude is reduced by $50$ percent, and $\Delta H$ is the half width of the layer. Fig.~\ref{fig:thin_layer} (blue curve) shows the radial decay profile when $H=1000$ km and $\Delta H = 100$ km. The flow speed is reduced from about $90$ percent of its surface value to $10$ percent over a distance of $200$ km. We compare this baroclinic zonal flow profile to a barotropic one where the observed surface winds extend $1000$ km deep (Fig.~\ref{fig:thin_layer}, orange line). For each case, we compute the odd zonal gravitational harmonics using the thermal wind equation (Fig.~\ref{fig:thin_layer}, table). The density perturbation within the thin layer mostly affects the value of $J_{3}$: the value for the baroclinic flow is about $100$ times smaller than the value for the barotropic flow. In addition to the calculation presented here, \citeA{galanti20} and \citeA{dietrich21} have applied decay functions that are similar to Eq.~\ref{eq:tanh} and reported $J_{3}$ values that are close to Juno-measured value of $-4.50 \pm 0.33 \times 10^{-8}$. This suggests that the $J_{3}$ values produced by the barotropic model could be reduced by accounting for the decay of the flow below the truncation depth. Future dynamical models, however, are needed to assess the amplitude of this effect. 

\subsection{Comparison of Barotropic and Baroclinic Solutions} 
\label{sec:compare}

Baroclinic zonal flow profiles that are solutions of the thermal wind equation and match the odd gravity harmonics derived from Juno measurements typically extend 3000 km deep \cite{kaspi18, kaspi20}. These baroclinic solutions are deeper than the $\sim 1000$ km barotropic solutions presented in this paper. Here, we show this difference in depth can be explained by the relationship between the background density and the zonal flow in the term, $\frac{\partial}{\partial z} \qty(\rho_{0}u_{\phi})$, in the thermal wind equation (Eq.~\ref{eq:twe}).  

\begin{figure}[t]
  \begin{adjustbox}{center}
    \includegraphics[width = 42pc]{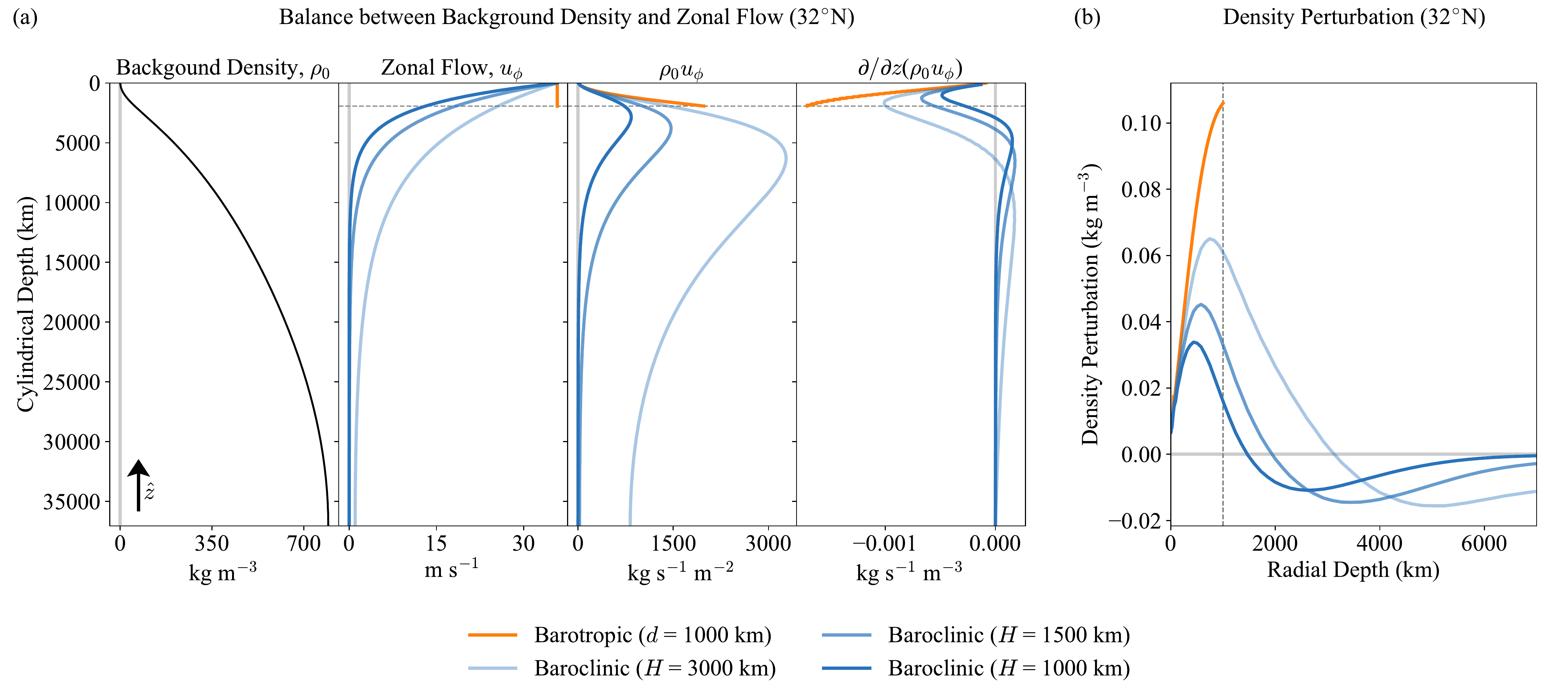}
  \end{adjustbox}
  \caption{We compare a barotropic solution of the thermal wind equation (Eq.~\ref{eq:twe}) where the observed surface winds extend into the interior to a radial depth of 1000 km to baroclinic solutions where the observed surface winds decay exponentially with different scale heights, $H$. (a) Components of the term, $\frac{\partial}{\partial z}\qty(\rho_{0}u_{\phi})$, as a function of cylindrical depth for a jet located at $32^{\circ}$N. (b) The density perturbation as a function of radial depth for a jet located at $32^{\circ}$N.}
  \label{fig:compare}
\end{figure}

Let us consider a jet in a barotropic zonal flow profile (Fig.~\ref{fig:compare}a, orange curve). As we move from the surface into the interior along the $z$-axis, the background density increases while the flow speed remains constant. The product, $\rho_{0}u_{\phi}$, increases with depth, reflecting the increase in the background density, and hence its derivative with respect to $z$ has only one sign in each hemisphere. When solving for the density perturbation, we integrate over the values of $\frac{\partial}{\partial z}\qty(\rho_{0}u_{\phi})$ along the $\theta$-direction. The net result is a density perturbation that is nonnegative everywhere (Fig.~\ref{fig:compare}b, orange curve). Since barotropic zonal flow is associated with a density perturbation of only one sign, deeper flows necessarily produce larger zonal gravitational harmonics. Thus, if the barotropic zonal flows exceed $\sim 1000$ km, they will produce odd zonal gravitational harmonics that are larger in magnitude than the Juno-derived values. 
 
Now, we consider a jet in a baroclinic zonal flow profile where the observed surface zonal winds decay exponentially with radial depth (Fig.~\ref{fig:compare}a, blue curves). As we move from the surface into the interior along the $z$-axis, the background density increases while the zonal flow speed decreases. The product, $\rho_{0}u_{\phi}$, initially increases, while the increase in background density overwhelms the decrease in flow speed, but later starts to decrease, as the flow speed approaches zero. Since $\rho_{0}u_{\phi}$ increases and then decreases with depth below the surface, its $z$-derivative, $\frac{\partial}{\partial z}\qty(\rho_{0} u_{\phi})$, changes signs. When we integrate over the $\frac{\partial}{\partial z}\qty(\rho_{0}u_{\phi})$ values, the resulting dynamical density field has a positive anomaly on top of a negative anomaly (Fig.~\ref{fig:compare}b, blue curves). Since the negative density anomaly partially cancels the gravitational signal of the positive density anomaly, baroclinic zonal flows need to extend beyond $\sim1000$ km in order to produce odd zonal gravitational harmonics that are consistent with the Juno values.

For all of the calculations presented in Section \ref{sec:results}, we neglected the dynamic self-gravitational term when solving for the flow-induced density perturbation. Including the dynamic self-gravity while assuming spherical background state and a rigid outer boundary would change our calculated $J_{3}$ values by about 30 percent \cite{wicht19}. Although this might be viewed as a first-order effect, we found that $J_{3}$ would also have first-order contributions from smooth mid/high latitude zonal flows. The shift in $J_{3}$ caused by the dynamic self-gravity could be offset by adjusting the structure and/or depth of the smooth mid/high latitude zonal flows. The dynamic self-gravity has less of an impact on the higher order odd zonal gravitational harmonics. It would change our calculated $J_{5}$, $J_{7}$, and $J_{9}$ values by about $10$ percent \cite{wicht19}, which would not alter our conclusions.

\section{Summary and Discussion} 
\label{sec:end}

The Juno gravitational measurements allow us to test physically motivated models of deep zonal flow. In this paper, we investigated a model in which Jupiter's deep atmospheric flows are barotropic, or invariant along the direction of the axis of rotation, until truncated at depth by some dynamical process (e.g., Reynolds stress, Lorentz or viscous force). We calculated the gravitational perturbations associated with the $z$-invariant part of the flows using the thermal wind equation and, from this, obtained the odd zonal gravitational harmonics ($J_{3}$, $J_{5}$, $J_{7}$, $J_{9}$). We found that $1244-1253$ km deep barotropic zonal flows involving the NOEJ and SOEJ could produce the Juno $J_{5}$, $J_{7}$, and $J_{9}$ values, but were not able to match the measured $J_{3}$ value. By combining the NOEJ and SOEJ with smooth mid/high latitude barotropic zonal flows, all of the Juno odd zonal gravitational harmonics could be explained.

While our analysis focused on non-equatorial flows, it is also worth considering possible structures of the equatorial flows within a barotropic model. The winds within the equatorial band ($14.0^{\circ}$S-$13.5^{\circ}$N) are strikingly symmetric. One potential explanation for this symmetry is that the two hemispheres are barotropically connected, representing Taylor-Proudman cylinders. Equatorial zonal flows with this structure are routinely produced in computer simulations modelling the deep convective outer layers of gas giant planets \cite{christensen01,aurnou01,heimpel05,jones09, gastine14z,yadav20_hex}. This explanation suggests that these flows extend to a depth of about $1800$ km at the equator. The symmetry of the equatorial zonal winds, however, does not necessarily require that they penetrate to such a depth. Nearly symmetrical equatorial flows are observed in other settings, such as in Earth's and Venus's atmospheres, where the flow is confined to a thin weather layer \cite{horinouchi17}.

As Juno continues to collect data during the extended mission, we can expect to further refine our barotropic solutions, especially in the mid/high latitudes. Juno's perijove will occur near $30^{\circ}$N at the start of the extended mission and migrate northward by about one degree with each additional orbit. As a result, Juno will become more sensitive to the gravitational field in the northern hemisphere. Tighter constraints on the gravitational field in this region could help us distinguish between different mid/high latitude zonal flow structures.

  Our key finding is that $\sim 1000$ km deep barotropic zonal flows are consistent with the Juno gravitational data. The question that then arises is: what dynamical processes could act to truncate the zonal flow? This question can be approached from two perspectives. The perspective usually considered in the literature is a top-down approach where it is assumed that the surface winds extend into the interior until they are disrupted at depth by a dynamical process, such as magnetic breaking or a stably stratified layer \cite{christensen20,gastine2021stable}. Alternatively, one could consider a bottom-up approach where dynamical processes excite the flow at depth and allow it to persist up to the weather layer  \cite{busse76,christensen01,aurnou01,heimpel05}. There are a few possible dynamical processes that could operate at a depth of $\sim 1000$ km for our barotropic model. One possibility is that the barotropic zonal flows are affected by a stably stratified layer, such as a rock cloud layer or a deep radiative zone \cite{markham18, guillot05}. Another possibility is that the $\sim 1000$ km deep flow is influenced by the magnetic field. While these flows are too shallow to experience direct magnetic breaking by the Lorentz force, it is possible that convective plumes originating in the dynamo region may penetrate into the overlying atmosphere and act to disturb the coherent turbulent processes needed to sustain zonal flows \cite{yadav20}. Our results therefore strongly advocate for further dynamical investigations of deep atmospheric zonal flows interacting with stably stratified layers and the dynamo region.

%% ------------------------------------------------------------------------ %%
 
%  ACKNOWLEDGMENTS
%
%% ------------------------------------------------------------------------ %%

\acknowledgments

This work is supported by the NASA Juno Mission. The Juno-derived zonal gravitational harmonics and their error estimates can be found in the Supporting Information for \citeNP{durante20}. The surface zonal wind profiles and the interior density model used in this study are available on Harvard Dataverse \cite{kulowski_bt_data}.

%% ------------------------------------------------------------------------ %%
%
% APPENDIX
% 
%% ------------------------------------------------------------------------ %%

\appendix
\section{Gravitational Signal of the Equatorial Jet}
\label{sec:equ_jet}

\begin{figure}[t]
  \begin{adjustbox}{center}
      \includegraphics[width=30pc]{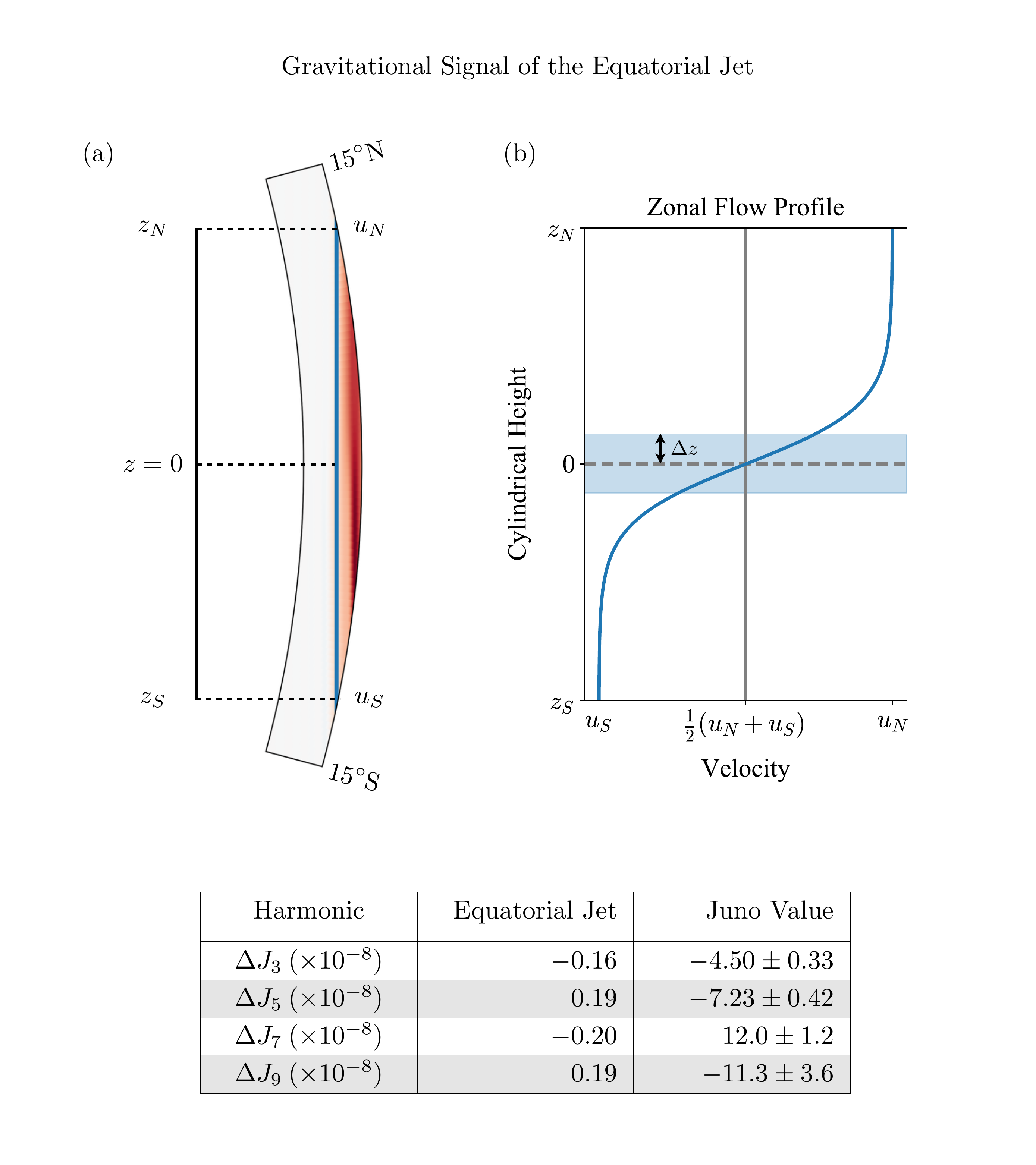}
    \end{adjustbox}
    \caption{(a) Equatorial zonal flow profile generated by connecting the northern and southern surface wind values, $u_{N}$ and $u_{S}$, using a sigmoid function (see Eq.~\ref{eq:sigmoid}). (b) Flow velocity as a function of $z$ for the blue highlighted cross-section shown in panel (a). The odd zonal gravitational harmonics produced by the flow are tabulated and compared to the Juno-derived values.} 
      \label{fig:jns_equ}
    \end{figure}

In this section, we calculate the maximum antisymmetric gravitational signal produced by the observed equatorial jet. We extend the observed surface winds between $14.0^{\circ}$S$-13.5^{\circ}$N barotropically to the equatorial plane. Due to the north-south asymmetries in the surface winds, we apply a smoothing function along the $z$-direction so that the flow at the equator equals the mean of the two surface values. This flow profile can be expressed using a sigmoid function 
\begin{align}
  u_{\phi}\qty(s, z) = \frac{u_{N}\qty(s) - u_{S}\qty(s)}{1 + \exp\qty(-\frac{z}{\Delta z})} + u_{S}\qty(s)
  \label{eq:sigmoid}
\end{align}
where $\qty(s, z)$ are cylindrical coordinates, $u_{N}$ and $u_{S}$ are the surface wind velocities in the northern and southern hemispheres, respectively, and $\Delta z$ is the half-width of the sigmoid function. The zonal flow profile when $\Delta z = 1500$ km is shown in Fig.~\ref{fig:jns_equ}a. The flow velocity as a function of $z$ is illustrated in Fig.~\ref{fig:jns_equ}b. The table in Fig.~\ref{fig:jns_equ} contains the low-degree odd zonal gravitational harmonics produced by this flow. At most, the equatorial jet produces odd harmonics that are one to two orders of magnitude smaller than the Juno-derived values. Thus, when analyzing Jupiter's antisymmetric gravitational field, the contribution from the equatorial jet can be neglected.

%% ------------------------------------------------------------------------ %%
%
% References and Citations
% 
%% ------------------------------------------------------------------------ %%

\bibliography{references}

\end{document}